\documentclass[12pt,oneside,letterpaper]{article}

\usepackage{graphicx,amsmath,amsfonts,amscd,amssymb,dcolumn,amsthm,textcomp,slashed,xspace,pgffor,tikz,latexsym,mathtools}
\usepackage[colorlinks,citecolor=blue,urlcolor=blue,linkcolor=blue]{hyperref}
\usepackage[T1]{fontenc}
\usepackage[utf8]{inputenc}
\usepackage{cite}
\usepackage{float}
\usepackage{xfrac}
\usepackage{footnote}
\usepackage{setspace}
\usepackage{caption}
\usepackage{url}
\usepackage{pifont}
\usepackage[final]{pdfpages}
\usepackage[margin=0.92in]{geometry}
\usepackage{indentfirst}
\usepackage{xcolor}
\usepackage{mathrsfs}
\usepackage[mathcal]{eucal}
\usepackage[hang]{footmisc} 
\usepackage{tcolorbox}

\tcbset{boxrule=0pt,colback=yellow,arc=0pt,auto outer arc,left=2pt,right=2pt,boxsep=2pt}

\numberwithin{equation}{section}

\theoremstyle{plain}

\begin{document}

\title{\textbf{Constrained gauge-gravity duality in three and four dimensions}}

\author{\textbf{T.~S.~Assimos}\thanks{thiago.assimos@gmail.com}\ , \textbf{R.~F.~Sobreiro}\thanks{rodrigo\_sobreiro@id.uff.br}\\\\
\textit{{\small UFF - Universidade Federal Fluminense, Instituto de Física,}}\\
\textit{{\small Campus da Praia Vermelha, Av. Litorânea s/n, 24210-346,}}\\
\textit{{\small Niterói, RJ, Brasil.}}}
\date{}
\maketitle
 
\begin{abstract}
The equivalence between Chern-Simons and Einstein-Hilbert actions in three dimensions established by A.~Achúcarro and P.~K.~Townsend (1986) and E.~Witten (1988) is generalized to the off-shell case. The technique is also generalized to the Yang-Mills action in four dimensions displaying de Sitter gauge symmetry. It is shown that, in both cases, we can directly identify a gravity action while the gauge symmetry can generate spacetime local isometries as well as diffeomorphisms. The price we pay for working in an off-shell scenario is that specific geometric constraints are needed. These constraints can be identified with foliations of spacetime. The special case of spacelike leafs evolving in time is studied. Finally, the whole set up is analyzed under fiber bundle theory. In this analysis we show that a traditional gauge theory, where the gauge field does not influence in spacetime dynamics, can be (for specific cases) consistently mapped into a gravity theory in the first order formalism.
\end{abstract}

\maketitle
\newpage
\tableofcontents
\newpage

\section{Introduction}

In\cite{Achucarro:1987vz}, A.~Achúcarro, P.~K.~Townsend and later E.~Witten \cite{Witten:1988hc} were able to show that Chern-Simons (CS) theory for the de Sitter and Poincaré groups are equivalent to three dimensional gravity with and without cosmological constant, respectively. In\cite{Achucarro:1987vz} the equivalence is discussed under a supersymmetric scenario while in\cite{Witten:1988hc} the non-supersymmetric scenario is also studied. Moreover, these results strongly depend on the validity of the classical field equations (on-shell cases). Essentially, the mapping between a CS theory and a gravity one consists on the equivalence between, for instance, the groups $SU(2)$ and $SO(3)$. Furthermore, an important feature is that the gauge symmetry contains local Lorentz isometries as well as diffeomorphisms, provided the validity of the classical field equations. In the case of the Poincaré group $ISO(3)$, the resulting gravity is the pure Einstein-Hilbert (EH) action while for the de Sitter group $SO(4)$ the resulting gravity action contains also the cosmological constant term. For a complete review on Chern-Simons theories and three-dimensional gravity we refer to \cite{Carlip:1998uc,Zanelli:2005sa,Marino:2005th,Zanelli:2012px,Hassaine:2016amq} and references therein.

It is possible to look at Gauge-Gravity Equivalence (GGE)\cite{Witten:1988hc} discussed in the previous paragraph as a mechanism to generate an effective gravity theory from a traditional gauge theory in flat spacetime\footnote{Throughout this work the term spacetime will always be associated with a Hausdorff differentiable manifold (or simply manifold).}. Essentially, we can state that the CS action for the group $ISO(3)\equiv SU(2)\times\mathbb{R}^3$ (or $SO(4)\equiv SU(2)\times S(3)$, where $S(3)$ describes gauge pseudo-translations. See Sec.~\ref{ads3d}.) is only a gauge theory, where the gauge group has no relation with spacetime dynamics. The spacetime itself can be taken as a flat manifold (Euclidean or Minkowskian), but this is not a requirement, only a simplification. Then, by making use of the homomorphism\footnote{The $SU(2)$ group is a double cover of $SO(3)$ and there is a $2\longmapsto1$ surjective homomorphism from $SU(2)$ to $SO(3)$.} $SU(2)\longmapsto SO(3)$, the gauge field can be mapped into the spin-connection and the dreibein. The sectors $\mathbb{R}^3$ and $S(3)$ in each case are identified with diffeomorphisms. The result is that the original spacetime is deformed into an effective manifold. For the general idea, see for instance\cite{Obukhov:1998gx,MacDowell:1977jt,Stelle:1979aj,Pagels:1983pq,Gotzes:1989wn,Tresguerres:2008jf,Tseytlin:1981nu,Mielke:2011zza,Sobreiro:2011hb,Sobreiro:2012iv,Assimos:2013eua}. Formally, this mechanism can be understood as a bundle map\cite{Kobayashi:1963fg,Nash:1983cq,Daniel:1979ez,Nakahara:2003nw} between a gauge principal bundle and a coframe bundle (See Sec.~\ref{fb}). Intuitively, one can think of the gauge fields as being "absorbed" by the spacetime.

In this work, besides the reinterpretation of the GGE\cite{Witten:1988hc} above described, we develop two main generalizations. First, we generalize such equivalence to the case where the field equations are no longer required to be obeyed (off-shell generalization). The main difference is that we have to consider a pair of constraints on the curvature and torsion 2-forms. Following \cite{Witten:1988hc}, we study the cases of the Poincaré ($ISO(3)$) and de Sitter ($SO(4)$) groups in three dimensions. Second, we generalize the GGE to four dimensions where we consider de Sitter group ($SO(5)$) as the gauge group of a Yang-Mills theory and employ the same ideas of\cite{Witten:1988hc} under the scenario described in the previous paragraph. We show that the gauge group can generate local Lorentz isometries as well as diffeomorphisms. The latter are originated from the coset $S(4)$ just like diffeomorphisms emerge from the coset $S(3)$ in the three-dimensional de Sitter case. Nevertheless, constraints are also demanded. We remark that, with little careful, everything can be systematically generalized to the general gauge group $SO(m,n)$ with $m+n=5$ in four dimensions. Nevertheless, for simplicity, we restrict ourselves to the case $m=0$, i.e., $SO(5)$. 

An important exigence of the mechanism is that a mass parameter is always required. This is evident if we realize that the gauge field is mass dimensionful while the vielbein is dimensionless. Thus, without a mass parameter, the identification between the gauge field and the vielbein may never be realized. In three dimensions, the coupling parameter is already dimensionful. Thus, there is a mass scale inherent to the theory. In contrast, in four dimensions, the coupling parameter is dimensionless. Therefore, it is necessary to impose the existence of a mass parameter to the theory. Fortunately, it is widely known that mass parameters are plenty in Yang-Mills theories\cite{Gribov:1977wm,Sobreiro:2005ec,Dudal:2005na,Dudal:2011gd,Pereira:2013aza,Capri:2015ixa,Capri:2016aqq,Capri:2016aif,Huber:2009tx,Cucchieri:2011ig,Cucchieri:2016jwg,Aguilar:2017dco}. On the other hand, in most models where gravity is generated from gauge theories\cite{MacDowell:1977jt,Stelle:1979aj,Pagels:1983pq,Gotzes:1989wn,Tresguerres:2008jf,Tseytlin:1981nu,Mielke:2011zza,Sobreiro:2011hb,Sobreiro:2012iv,Assimos:2013eua,Sobreiro:2016fks}, a symmetry breaking, or/and an Inönü-Wigner contraction\cite{Inonu:1953sp}, is required. In the present approach, no breaking mechanism is needed although a mass parameter existence is mandatory.

Furthermore, we also discuss the geometrical meaning of these constraints and argue that they are equivalent to spacetime foliations. The specific case of the spacelike leafs evolving in time is then discussed. We find that, for the constraints to be fulfilled at the off-shell cases, the diffeomorphism symmetry must be broken to a smaller diffeomorphic group.

Finally, we provide a study of the GGE we develop in terms of fiber bundle theory \cite{Kobayashi:1963fg,Nash:1983cq,Daniel:1979ez,Nakahara:2003nw}. A gauge theory is defined over a principal bundle whose structure group is the gauge group and does nor affect the base space (spacetime). Gravity, on the other hand, is constructed over a coframe bundle. The coframe bundle is also a principal bundle but with the special property that the structure group (gauge group) is identified with the base space isometries (local isometries of the spacetime). Essentially, we discuss how to identify a gauge principal bundle with a coframe bundle. 

This work is organized as follows: In Sec.~\ref{3d} we review the three-dimensional case as first realized by Witten and generalize it to the off-shell case. In Sec.~\ref{4d}, the four-dimensional case for the Yang-Mills action is discussed. In Sec.~\ref{adm} the constraints are discussed from viewpoint of a specific foliation where spacelike leafs evolve in time. Then, in Sec.~\ref{fb} we formally discuss how to reinterpret Witten's GGE in terms of fiber bundles. Finally, the conclusions and perspectives are displayed in Sec.~\ref{concl}.

\section{Constrained gravity in three dimensions}\label{3d}

In this section we stick to three dimensions. We review and generalize the non-supersymmetric situation discussed in\cite{Witten:1988hc}.

\subsection{Poincaré gauge symmetry and gravity}\label{eh3d}

The Chern-Simons action is given by
\begin{equation}\label{cs1}
S_{_{CS}} = \dfrac{\varsigma}{4\pi}\mathrm{Tr}\int \left(Y dY + \frac{2}{3}YYY\right)\;,
\end{equation}
where $\varsigma$ is the Chern-Simons coupling and $Y$ is the 1-form connection for the Poincaré group in the representation $ISO(3)\equiv SU(2)\times\mathbb{R}^3$, which is the gauge group that leaves the action \eqref{cs1} invariant. In this representation, the following conventions for the algebra are assumed,
\begin{eqnarray}\label{alg1}
\left[J_{i},J_{j}\right]&=&{\epsilon_{ijk}}J^{k}\;,\nonumber\\
\left[J_{i},P_{j}\right]&=&{\epsilon_{ijk}}P^{k}\;, \nonumber \\ 
\left[P_{i},P_{j}\right]&=& 0\;,
\end{eqnarray}
and for the invariant quadratic forms it is considered that
\begin{eqnarray}\label{iqf1}
\mathrm{Tr}(J^{i}J^{j})&=&0\;,\nonumber\\
\mathrm{Tr}(P^{i}P^{j})&=&0\;,\nonumber\\
\mathrm{Tr}(J^{i}P^{j})&=&\delta^{ij}\;, 
\end{eqnarray}
where small Latin indices vary as $\{0,1,2\}$. Decomposing the gauge field accordingly,
\begin{equation}\label{dec1}
Y=A^{i}J_{i}+\theta^{i}P_{i}\;,
\end{equation}
the action \eqref{cs1} can be rewritten as
\begin{equation}\label{act1}
S_{_{CS}}=\dfrac{\varsigma}{2\pi}\int\theta^{i}F_{i}\;, 
\end{equation}
where $F^{i}=dA^{i}+\frac{1}{2}\epsilon^{ijk}A_{j}A_{k}$ is the 2-form field strength associated to the $su(2)$ sector of the algebra. 

The transition to the $ISO(3)\equiv SO(3)\times \mathbb{R}^3$ representation can be performed by the following identifications
\begin{eqnarray}\label{id1}
A^i&\longmapsto&-\frac{1}{2}\epsilon^{ijk}\omega_{jk}\;,\nonumber \\
\theta^i&\longmapsto&\kappa^{2} e^{i}\;,
\end{eqnarray}
where $\kappa$ is a parameter with $1/2$ mass dimension associated with the coupling parameter while $e^{i}$ and $\omega^{ij}$ are recognized as the dreibein and the the spin-connection, respectively. Similarly to the relations \eqref{id1}, the algebra \eqref{alg1} can also be transformed into an $SO(3)\times\mathbb{R}^3$ algebra through 
\begin{equation}\label{dual1}
J^{i}=\frac{1}{2}\epsilon^{i}_{\phantom{i}jk}J^{jk}\;,
\end{equation}
providing
\begin{eqnarray}\label{alg2}
\left[J^{ij},J^{kl}\right]&=&-\frac{1}{2}\left[\left(\eta^{ik}J^{jl}+\eta^{jl}J^{ik}\right)-\left(\eta^{il}J^{jk}+\eta^{jk}J^{il}\right)\right]\;,\nonumber\\
\left[P^i,P^j\right]&=&0\;, \nonumber\\
\left[J^{ij},P^k\right]&=&\frac{1}{2}\left(\eta^{ik}J^j-\eta^{jk}J^i\right)\;. 
\end{eqnarray}
The substitution of \eqref{id1} in \eqref{act1} results in 
\begin{equation}\label{eh1}
S_{_{CS}}\longmapsto S_{_{EH3d}}=-\frac{1}{16\pi G}\int\epsilon_{ijk}e^{i}R^{jk}\;,
\end{equation}
where $R^{ij}=d\omega^{ij}+{\omega^i} _k\omega^{kj}$ is the curvature 2-form. The "Newton's constant"\footnote{The quotation marks are employed because gravity theory in three dimensions does not have a Newtonian limit. Nevertheless, since $G$ appear in front of the three-dimensional Einstein-Hilbert action, it can be associated with the four-dimensional Newton's constant by analogy.} is determined by $G=\dfrac{1}{8\varsigma\kappa^2}$. Needless to say, the action \eqref{eh1} is the three dimensional Einstein-Hilbert (EH) action. The respective field equations are simply $R=T=0$, where $T=\widetilde{D}e$ is the torsion 2-form and $\widetilde{D}=d+\omega$ is the covariant derivative with respect to the $SO(3)$ sector.

Let us take a look at the gauge transformations. Originally (for the representation $SU(2)\times\mathbb{R}^3$), they are given by
\begin{eqnarray}\label{gt1}
A^{i}&\longmapsto&A^{i}+D\alpha^{i}\;, \nonumber \\
\theta^{i}&\longmapsto&\theta^{i}+D\xi^{i}+\epsilon^i_{\phantom{i}jk}\alpha^{j}\theta^{k}\;, 
\end{eqnarray}
where $\zeta=\alpha^{i}J_{i}+\xi^{i}P_{i}$ is the full infinitesimal gauge parameter and $D=d+A$ is the covariant derivative. Using \eqref{id1}, it is easy to show that these transformations can be rewritten as
\begin{eqnarray}\label{gt2}
\omega^{ij}&\longmapsto&\omega^{ij}+\widetilde{D}\alpha^{ij}\;, \nonumber \\
e^{i}&\longmapsto&e^{i}+{\widetilde{D}}\rho^{i}+{\alpha^i}_je^j\;,
\end{eqnarray}
where we have employed the extra identifications
\begin{eqnarray}\label{id2}
\alpha^{ij}&=&-\epsilon^{ijk}\alpha_{k}\;,\nonumber\\
\rho^i&=&\dfrac{1}{\kappa^{2}}\xi^i\;,
\end{eqnarray}
characterizing the map $SU(2)\longmapsto SO(3)$.

The main step to achieve a complete gravity theory, according to\cite{Witten:1988hc}, is to associate the $SO(3)$ sector with local Lorentz transformations and the sector $\mathbb{R}^3$ with diffeomorphisms. In fact, diffeomorphisms can be obtained from the Lie derivative\cite{Kobayashi:1963fg,Wald:1984rg}, 
\begin{equation}\label{diff1}
\mathfrak{L}_v\cdot=i_v(d\cdot)+d(i_v\cdot)\;,
\end{equation}
where $i_v$ is the interior derivative, which reduces the form rank by one\footnote{The interior derivative acts as $i_v(dx^\nu):=v^\nu$. See, for instance, \cite{Bertlmann:1996xk}.}. The subindex $v$ characterizes an element in a vector space $\mathcal{V}(\mathbb{M}^3)$, where $\mathbb{M}^3$ is the spacetime manifold. Then, $v^\mu$ is identified with the diffeomorphism parameters. The Greek indices are world indices and vary as $\{0,1,2\}$. Therefore, it is quite simple to show that,
\begin{eqnarray}\label{diff2}
\mathfrak{L}_v\omega^{ij}&=& i_vR^{ij}-i_v(\omega^{i}_{\phantom{i}k}\omega^{kj})+d(i_v\omega^{ij})\;\;=\;\; i_vR^{ij}+\widetilde{D}(i_v\omega^{ij})\;,\nonumber \\
\mathfrak{L}_ve^{i} &=& i_vT^i-i_v({\omega^i}_ke^k) + d(i_ve^i)\;\;=\;\;i_vT^i+\widetilde{D}(i_ve^i)-i_v{\omega^i}_ke^k\;.
\end{eqnarray}
On the other hand, the gauge transformations \eqref{gt2}, when restricted to the $\mathbb{R}^3$ sector, read
\begin{eqnarray}\label{gt3}
\delta_{_{\mathbb{R}^3}}\omega^{ij}&=&0\;,\nonumber\\
\delta_{_{\mathbb{R}^3}}e^{i}&=&{\widetilde{D}}\rho^{i}\;.
\end{eqnarray}
The difference between transformations \eqref{diff2} and \eqref{gt3} provides
\begin{eqnarray}\label{isom1}
(\mathfrak{L}_v-\delta_{_{\mathbb{R}^3}})\omega^{ij}&=&i_vR^{ij}+\widetilde{D}(i_v\omega^{ij})\;,\nonumber\\
(\mathfrak{L}_v-\delta_{_{\mathbb{R}^3}})e^i &=& i_vT^i+\widetilde{D}\left(i_ve^i-\rho^i-(i_v{\omega^i}_k)\right)e^k\;,
\end{eqnarray}
which reduce to local Lorentz local transformations if we set
\begin{eqnarray}\label{rel1}
\alpha^{ij}&=&i_v\omega^{ij}\;,\nonumber\\
\rho^i&=&i_ve^i\;,
\end{eqnarray}
and
\begin{eqnarray}\label{rel2}
i_vR^{ij}&=&0\;,\nonumber \\
i_vT^i&=&0\;.
\end{eqnarray}

Equations \eqref{rel2} define constraints over the gravity that emerges from the Poincaré invariant Chern-Simons action \eqref{cs1}. For general geometries (specific non-vanishing curvature and torsion), the meaning of these equations are simple: given a geometry, the diffeomorphic vector $v^\mu$ must be invariant if parallel transported along an infinitesimal closed path. The possible solutions of these constraint are:
\begin{itemize}

\item [(i)] $R^{ij}= 0$; $T^i=0$, which is the on-shell case discussed in\cite{Witten:1988hc}. This case displays full diffeomorphism symmetry but the classical vacuum solutions must be imposed. We remark that, if matter is included then this situation does not rely on the on-shell case anymore.

\item [(ii)] $v^\nu = 0$; which means that diffeomorphism symmetry is entirely broken. Obviously, this is not an interesting situation if diffeomorphisms are desired as a property of the final result.

\item [(iii)] The most general case is to consider the constraints \eqref{rel2} in a situation different from (i) and (ii). In fact, these constraints formally define a spacetime regular foliation\cite{Dufour:2005th,Lavau:2018th} with partial breaking of the diffeomorphisms, i.e., an intermediate situation between (i) and (ii). Hence, in the present case, the full diffeomorphism symmetry is then broken to a smaller group,
\end{itemize}
\begin{equation}\label{sub1}
\mathrm{Diff}(2)_{\mathrm{fol}}=\{v\in \mathrm{Diff}(3)\;\big|\;i_vR^{ij}=0\;\;;\;\;i_vT^i=0\}\;.
\end{equation}
Clearly, situations (i) and (ii) are extreme limits of situation (iii).

\subsection{de Sitter gauge symmetry and gravity}\label{ads3d}

The second case considered in\cite{Witten:1988hc} is the generalization of the previous approach to the action \eqref{cs1} for the gauge group $SO(4) \equiv SU(2)\times S(3)$, where $S(3)$ are pseudo-translations. The translational sector of the algebra \eqref{alg1} is modified by $[P_{i},P_{j}]=\epsilon_{ijk} J^{k}$ while the invariant quadratic forms \eqref{iqf1} are exactly the same. Employing again decomposition \eqref{dec1}, but with the algebra of the $SO(4)$ group, to the Chern-Simons action \eqref{cs1}, we find
\begin{equation}\label{cs2}
S_{_{CS}}=\dfrac{\varsigma}{2\pi}\int\theta^{i}\left(F_{i}+\frac{1}{6}\epsilon_{ijk}\theta^{j}\theta^{k}\right)\;.
\end{equation}
Obviously, the field $\theta$ now belongs to the pseudo-translational sector. Then, using the identifications \eqref{id1} again, the Chern-Simons action is mapped into
\begin{equation}\label{eh2}
S_{_{CS}}\longmapsto S_{_{grav}}=-\frac{1}{16\pi G}\int\epsilon_{ijk}e^{i}\left(R^{jk}-\frac{\Lambda^2}{3}e^{j} e^{k}\right)\;, 
\end{equation}
where $\Lambda^2=\kappa^4$ is identified with the cosmological constant. "Newton's constant" remains fixed as $G=\dfrac{1}{8\varsigma\kappa^2}$. Clearly, the action \eqref{eh2} is invariant under the orthogonal group $SO(4)$ in the representation $SO(3)\times S(3)$.

In order to encounter a possible diffeomorphic symmetry from the gauge symmetry, we write the original gauge transformations for the fields $A$ and $\theta$, namely,
\begin{eqnarray}\label{gt4}
A^{i}&\longmapsto&A^{i}+D\alpha^i+{\epsilon^i}_{jk}\theta^j\xi^k\;,\nonumber\\
\theta^{i}&\longmapsto&\theta^{i}+D\xi^{i}+\epsilon^{i}_{\phantom{i}jk}\theta^{k}\alpha^j\;. 
\end{eqnarray}
Applying \eqref{id1} and \eqref{id2}, we find
\begin{eqnarray}\label{gt5}
\omega^{ij}&\longmapsto&\omega^{ij}+\widetilde{D}\alpha^{ij}-\Lambda^2\left(e^i\rho^j-e^j\rho^i\right)\;, \nonumber \\
e^{i}&\longmapsto&e^{i}+{\widetilde{D}}\rho^{i}+\alpha^{i}_{\phantom{i}j}e^{j}\;. 
\end{eqnarray}
In this case, the gauge group changed to the representation $SO(4) \equiv SO(3) \times S(3)$. Again, we wish to identify $S(3)$ with diffeomorphisms. For that, we consider once again the relations \eqref{diff2}. The gauge transformations \eqref{gt5}, restricted to the $S(3)$ sector, read now
\begin{eqnarray}\label{gt6}
\delta_{_{S(3)}}\omega^{ij}&=&-\Lambda^2\left(e^i\rho^j-e^j\rho^i\right)\;,\nonumber\\
\delta_{_{S(3)}}e^{i}&=& {\widetilde{D}} \rho^i\;.
\end{eqnarray}
Thus,
\begin{eqnarray}\label{isom2}
(\mathfrak{L}_v-\delta_{_{S(3)}})\omega^{ij}&=&i_vR^{ij}+\widetilde{D}(i_v\omega^{ij})+\Lambda^2\left(e^i\rho^j-e^j\rho^i\right)\;,\nonumber\\
(\mathfrak{L}_v-\delta_{_{S(3)}})e^i&=&i_vT^i+\widetilde{D}(i_ve^i-\rho^i)-(i_v{\omega^i}_k)e^k\;.
\end{eqnarray}
The relations \eqref{isom2} reduce to local Lorentz local transformations if, and only if, we set relations \eqref{rel1} and 
\begin{eqnarray}\label{rel4}
i_v\left(R^{ij}-\Lambda^2e^ie^j\right)&=&0\;,\nonumber\\
i_vT^i&=&0\;.
\end{eqnarray}
The difference between the constraints \eqref{rel2} and \eqref{rel4} is the presence of the cosmological constant term. In fact, the first of the constraints \eqref{rel4} can be interpreted as a condition over the curvature when filtered by the vacuum solution. Under this interpretation, constraints \eqref{rel4} is no different from \eqref{rel2} where the vacuum solution is null curvature and torsion.

From \eqref{rel4} we can see that $R^{ij}=\Lambda^ 2e^ie^j$ and $T^i=0$ consist on the on-shell solution, as discussed in\cite{Witten:1988hc}. This is the analogous case of (i) in the previous section. If $v^\nu=0$, the situation (ii) of the previous section is achieved, corresponding to the full breaking of diffeomorphisms. Again, the intermediate case (i.e., the case analogous to (iii)) would be a specific foliation of spacetime, providing a diffeomorphisms subgroup characterized by
\begin{equation}\label{sub1a}
\mathrm{Diff}(2)_{\mathrm{fol}}=\{v\in \mathrm{Diff}(3)\;\big|\;i_v\left(R^{ij}-\Lambda^2e^ie^j\right)=0\;\;;\;\;i_vT^i=0\}\;.
\end{equation}
The geometric interpretation of the constraints \eqref{rel4} are now different: the parallel transport of $v^\mu$ along an infinitesimal path will start and end at the same point (as before) but will not remain invariant; it will be modified according to a de Sitter curvature $\Lambda^2e^ie^j$.

\section{Constrained gravity in four dimensions}\label{4d}

We consider now a gauge theory for the group $SO(5)$ over a four-dimensional manifold with Euclidean signature. The gauge group can be decomposed according to $SO(5)\equiv SO(4)\times S(4)$. The respective algebra splits as
\begin{eqnarray}\label{alg3}
\left[J^{ab},J^{cd}\right]&=&-\frac{1}{2}\left[\left(\eta^{ac}J^{bd}+\eta^{bd}J^{ac}\right)-\left(\eta^{ad}J^{bc}+\eta^{bc}J^{ad}\right)\right]\;,\nonumber\\
\left[J^a,J^b\right]&=&-\frac{1}{2}J^{ab}\;,\nonumber\\
\left[J^{ab},J^c\right]&=&\frac{1}{2}\left(\eta^{ac}J^b-\eta^{bc}J^a\right)\;,
\end{eqnarray}
where $J^{ab}$ are anti-hermitian generators. Now, small Latin indices vary as $\{0,1,2,3\}$. Accordingly, the gauge connection also splits through $Y=A^a_{\phantom{a}b}J_a^{\phantom{a}b}+\theta^aJ_a$. Consequently, the curvature 2-form can be written as
\begin{equation}\label{fieldstr}
F=\left(\varSigma^{a}_{\phantom{a}b}-\frac{1}{4}\theta^{a}\theta_{b}\right)J_{a}^{\phantom{a}b}+\varPhi^{a}J_{a}\;,
\end{equation}
where $\Sigma^{a}_{\phantom{a}b}=dA^a_{\phantom{a}b}+A^a_{\phantom{a}c}A^c_{\phantom{c}b}$ and $\Phi^{a}=d\theta^{a}+ A^a_{\phantom{a}b}\theta^{b}$.

If we consider pure Yang-Mills action as the starting theory, it decomposes as 
\begin{eqnarray}\label{ym1}
S_{_{YM}}&=&\frac{1}{2\kappa^2}\int F^\mathcal{A}_{\phantom{\mathcal{A}}\mathcal{B}}{*} F_\mathcal{A}^{\phantom{\mathcal{A}}\mathcal{B}}\nonumber\\
&=&\frac{1}{2\kappa^2}\int\left[\Sigma^a_{\phantom{a}b}{*}\Sigma_a^{\phantom{a}b}+\frac{1}{2}\Phi^{a}{*}\Phi_{a}-\frac{1}{2}\Sigma^a_{\phantom{a}b}{*}(\theta_a\theta^b)+\frac{1}{16}\theta^a\theta_b{*}(\theta_a\theta^b)\right]\;,
\end{eqnarray}
where calligraphic index vary as $\{0,1,2,3,4\}$. Just like the three-dimensional case, $\kappa$ stands for the coupling parameter. The difference is that in four dimensions $\kappa$ is dimensionless, $[\kappa]=0$.

Following the ideas developed in the previous sections, it is possible to infer the following map between the gauge field and the first order gravity variables (see for instance\cite{Sobreiro:2011hb,Assimos:2013eua})
\begin{eqnarray}\label{id3}
A^{a}_{\phantom{a}b} &\longmapsto& \omega^{a}_{\phantom{a}b}\;,\nonumber\\
\theta^{a} &\longmapsto& \gamma e^{a}\;,
\end{eqnarray}
where $\gamma$ is a mass scale. This mass scale is required because $[\kappa]=0$ and $[\theta]=[A]=1$ while $[\omega]=1$ and $[e]=0$, provided $[\gamma]=1$. We remark that, in three dimensions, the coupling parameter already has mass dimension. In four-dimensions, there are no former mass parameters available. Thus, its existence must be assumed\footnote{In fact, such parameter can emerge from non-perturbative effects as condensates or the Gribov parameter, see for instance\cite{Gribov:1977wm,Sobreiro:2005ec,Dudal:2005na,Dudal:2011gd,Pereira:2013aza,Capri:2015ixa,Capri:2016aqq,Capri:2016aif} and references therein.}. Hence, the Yang-Mills action \eqref{ym1} assumes the form
\begin{equation}\label{ym2}
S_{_{4d(grav)}}=\frac{1}{16\pi G}\int\left[\frac{3}{2\Lambda^2}{R^a}_b\ast {R_a}^b+T^a\ast T_a-\frac{1}{2}\epsilon_{abcd}e^ae^b\left(R^{cd}-\frac{\Lambda^{2}}{6} e^ce^d\right)\right]\;,
\end{equation}
where $R^{a}_{\phantom{a}b}=d\omega^{a}_{\phantom{a}b}+\omega^{a}_{\phantom{a}c}\omega^{c}_{\phantom{c}b}$, $T^a = de^a+\omega^{a}_{\phantom{b}b}e^b$, $G=\dfrac{\kappa^2}{4\pi\gamma^2}$ and $\Lambda^2=\dfrac{3\gamma^2}{4}$ are recognized as the curvature and torsion 2-forms, the Newton and cosmological constants, respectively. The most simple nontrivial solution of the field equations with $T^a=0$ is unusual because it provides $R^{ab}_{\phantom{ab}\mu\nu}\in\mathbb{C}$.

The gauge transformations associated with the original $SO(5)$ theory are
\begin{eqnarray}\label{gt7}
{A}^a_{\phantom{a}b}&\longmapsto& {A}^a_{\phantom{a}b}+\nabla\alpha^a_{\phantom{a}b}-\frac{1}{4}\left(\theta^a\xi_b-\theta_b\xi^a\right)\;,\nonumber\\
\theta^a&\longmapsto&\theta^a+\nabla\xi^a+\alpha^a_{\phantom{a}b}\theta^b\;,
\end{eqnarray}
where $\zeta=\alpha^a_{\phantom{a}b}J_a^{\phantom{a}b}+\xi^aJ_a$ and $\nabla={d}+A$ is the covariant derivative with respect to the sector $SO(4)$. Using the identifications \eqref{id3}, it is easy to show that these transformations can be rewritten as
\begin{eqnarray}\label{gt8}
\omega^a_{\phantom{a}b} &\longmapsto&\omega^a_{\phantom{a}b}+\widetilde\nabla\alpha^a_{\phantom{a}b}-\frac{\Lambda^2}{3}\left(e^a\rho_b - e_b\rho^a\right)\;,\nonumber\\
e^a &\longmapsto&e^a+\widetilde\nabla\rho^a-\alpha^a_{\phantom{a}b}e^b\;,
\end{eqnarray}
with $\widetilde\nabla={d}+\omega$ and
\begin{equation}\label{id4}
\rho^a=\frac{1}{\gamma}\xi^a\;.
\end{equation}
If we select the $S(4)$ sector of the transformations, we achieve
\begin{eqnarray}\label{gt9}
\delta_{_{S(4)}}\omega^a_{\phantom{a}b} &=&-\frac{\Lambda^2}{3}\left(e^a\rho_b-e_b\rho^a\right)\;,\nonumber\\
\delta_{_{S(4)}} e^a &=& \widetilde\nabla\rho^a\;.
\end{eqnarray}

Again, we wish to obtain diffeomorphisms from the gauge group. This time, from the $S(4)$ part. Thus, following the prescription of the three dimensional case, we can obtain the diffeomorphic transformations of the fields from the Lie derivative, namely,
\begin{eqnarray}\label{diff3}
\mathfrak{L}_v\omega^{a}_{\phantom{a}b} &=&i_{v}R^a_{\phantom{a}b}-i_v(\omega^{a}_{\phantom{a}c}\omega^{c}_{\phantom{c}b})+ d(i_v\omega^{a}_{\phantom{a}b})\;\;=\;\; i_{v}R^a_{\phantom{a}b}+\widetilde{\nabla}(i_v\omega^{a}_{\phantom{a}b}),\nonumber \\
\mathfrak{L}_ve^{a} &=&i_{v}T^a-i_v({\omega^a}_ce^c) + d(i_ve^a)\;\;=\;\;i_{v}T^a+\widetilde{\nabla}(i_ve^a)-i_v{\omega^a}_ce^c\;.
\end{eqnarray}
Taking the difference between \eqref{gt9} and \eqref{diff3}, we obtain
\begin{eqnarray}\label{diff4}
(\mathfrak{L}_v-\delta_{_{S(4)}})\omega^{a}_{\phantom{a}b} &=&i_{v}R^a_{\phantom{a}b}+\widetilde{\nabla}(i_v\omega^{a}_{\phantom{a}b})+\frac{\Lambda^2}{3}\left(e^a\rho_b-e_b\rho^a\right)\;,\nonumber\\
(\mathfrak{L}_v-\delta_{_{S(4)}})e^a &=&i_{v}T^a+\widetilde{\nabla}(i_ve^a)-i_v{\omega^a}_ce^c-\widetilde\nabla\rho^a\;.
\end{eqnarray}
Assuming analogous relations to \eqref{rel1}, given by
\begin{eqnarray}\label{rel8}
{\alpha^a}_b&=&i_v{\omega^a}_b\;,\nonumber\\
\rho^a&=&i_ve^a\;,
\end{eqnarray}
and imposing the constraints
\begin{eqnarray}\label{rel9}
i_v\left(R^{ab}-\frac{\Lambda^2}{3}e^ae^b\right)&=&0\;,\nonumber\\
i_vT^a&=&0\;,
\end{eqnarray}
the $S(4)$ sector of the gauge algebra can be identified with diffeomorphisms. 

Differently from the three-dimensional cases we do not have the three particular possibilities of constraints discussed at the end of Sec.~\ref{ads3d}. Inhere there is no room for the on-shell case:
\begin{itemize}
\item [(i)] de Sitter curvature and vanishing torsion ($R^{ab}=\dfrac{\Lambda^2}{3}e^ae^b$ and $T^a=0$): This case displays full diffeomorphism. However, this is not the on-shell case. Instead, this geometry sets $S_{_{4d(grav)}}=0$.

\item [(ii)] $v^\nu = 0$; meaning that diffeomorphism symmetry is entirely broken and the geometry is totally free.

\item [(iii)] The most general case is the intermediate situation between (i) and (ii) where \eqref{rel9} still holds and the full diffeomorphism symmetry is then broken to a smaller diffeomorphic group,
\end{itemize}
\begin{equation}\label{sub1b}
\mathrm{Diff}(3)_{\mathrm{fol}}=\left\{v\in \mathrm{Diff}(4)\;\big|\;i_v\left(R^{ab}-\dfrac{\Lambda^2}{3}e^ae^b\right)=0\;\;;\;\;i_vT^a=0\right\}\;.
\end{equation}
The geometrical interpretation of the constraints \eqref{rel9} is similar to the de Sitter three-dimensional case.

\section{Analysis of the constraints in foliated spacetime}\label{adm}

Given the generic constraints in three and four dimensions (cases (iii)), namely \eqref{rel2}, \eqref{rel4} and \eqref{rel9}, it is our intent to verify their consistency. The analysis we perform in this section is quite general, but we rather be closer to Physics and think of a spacetime manifold is foliated as $\mathbb{M}^d=\mathbb{M}^{(d-1)}\times\mathbb{R}_t$, where $\mathbb{R}_t$ and $\mathbb{M}^{(d-1)}$ are the temporal and spatial parts, respectively. In particular, this choice is consistent with the Arnowitt-Deser-Misner (ADM) foliation formalism \cite{Arnowitt:1959ah,Misner:1974qy,Jacobson:1988yy,Rovelli:1991zi,Lusanna:1998pu,Mei:2007xk,Alexandrov:2014rta}. In this approach the dynamics of the gravitational field can be seen as the time evolution of these hypersurfaces along time. 

The relation between a $(d-1)$-dimensional foliation curvature $\mathcal{R}$, the $d$-dimensional curvature of the spacetime $R$ and the extrinsic curvature $K^I_{\phantom{I}\mu}$ of the foliation, is given by the Gauss equation
\begin{eqnarray}\label{gauss1}
\mathcal{R}^{IJ}_{\phantom{IJ}\mu\nu}&=&R^{\alpha\beta}_{\phantom{\alpha\beta}\mu\nu}e^I_{\phantom{I}\alpha}e^J_{\phantom{J}\beta}-K^I_{\phantom{I}\mu}K^{J}_{\phantom{J}\nu}+K^J_{\phantom{J}\mu}K^I_{\phantom{I}\nu}\;.
\end{eqnarray}
The spatial indices refer to coordinates on the leaf, i.e., $\{I,J\}=\{1,2,3,\ldots d-1\}$. The extrinsic curvature is defined with the vector normal to the leaf, $n^\mu$, as $K^I_{\phantom{I}\mu}=-e^I_{\phantom{I}\nu}\widetilde{D}_{\mu}n^{\nu}$. The decomposition of the $d$-dimensional torsion tensor $T$ in the leaf is given by
\begin{eqnarray}\label{gausstors}
\mathcal{T}^{I}_{\phantom{I}\mu\nu}&=&e^I_{\phantom{I}\alpha}T^{\alpha}_{\phantom{\alpha}\mu\nu}+\Omega^I_{\phantom{I}\mu\nu}\;,
\end{eqnarray}
where $\mathcal{T}$ is the projected torsion, i.e., the $(d-1)$-dimensional torsion of the leaf and the rest, in analogy to \eqref{gauss1}, is related to the extrinsic torsion. For simplicity and due to the fact that we will not make use of its explicit form we will refer to $\Omega^I_{\phantom{I}\mu\nu}$ simply as "extrinsic torsion".

\subsection{Three dimensions}

In the case of three dimensions we refer to the constraints \eqref{rel2} and \eqref{rel4} for Poincaré and de Sitter symmetries, respectively. For the sake of simplicity, let us consider some special cases before the general ones.

\subsubsection{Poincaré symmetry}\label{3dconst}

\begin{itemize}
 \item[(I)]{\textbf{On-shell torsion and general curvature}}
\end{itemize}
On-shell torsion in this case case means $T^i=0$ and general curvature means that $R^{ij}$ is arbitrary, except for the condition $i_vR^{ij}=0$. This is the usual Riemannian scenario of gravity. Thus, the full diffeomorphism symmetry is broken to a smaller group,
\begin{equation}\label{sub1dif}
\mathrm{Diff}(2)_{\mathbb{M}^2}=\{v\in \mathrm{Diff}(3)\;\big|\;\mathbb{M}^3=\mathbb{M}^2\times\mathbb{R}_t\;;\;\;T^i=0\;\;;\;\;i_vR^{ij}=0\}\;.
\end{equation}

To find the specific conditions over the diffeomorphisms parameters $v^\mu$, we start by multiplying the Gauss equation \eqref{gauss1} in three dimensions by $v^\mu$,
\begin{eqnarray}\label{gauss2}
\mathcal{R}^{IJ}_{\phantom{IJ}\mu\nu}v^\nu&=& e^I_{\phantom{I}\alpha}e^J_{\phantom{J}\beta}R^{\alpha\beta}_{\phantom{\alpha\beta}\mu\nu}v^\nu +\left(-K^I_{\phantom{I}\mu}K^{J}_{\phantom{J}\nu}+K^J_{\phantom{J}\mu}K^I_{\phantom{I}\nu}\right)v^\nu\;.
\end{eqnarray}
The first term of the \emph{rhs} in \eqref{gauss2} vanishes due to the constraint \eqref{rel2}, yielding
\begin{eqnarray}\label{gauss2a}
\mathcal{R}^{IJ}_{\phantom{IJ}\mu\nu}v^\nu&=&\left(-K^I_{\phantom{I}\mu}K^{J}_{\phantom{J}\nu}+K^J_{\phantom{J}\mu}K^I_{\phantom{I}\nu}\right)v^\nu\;.
\end{eqnarray}
Expression \eqref{gauss2a} expresses the behavior of the field $v^\mu$ when parallel transported along a spacelike curve (a circuit lying on the spatial leaf of the foliation). The corresponding constraint on $v^\mu$ can easily be found by rewriting \eqref{gauss2a} in components, providing
\begin{eqnarray}\label{gauss4}
\mathcal{R}^{IJ}_{\phantom{IJ}{P0}}v^P&=&0\;,
\end{eqnarray}
and
\begin{eqnarray}\label{gauss5}
\mathcal{R}^{IJ}_{\phantom{IJ}{P0}}v^0+\mathcal{R}^{IJ}_{\phantom{IJ}{PQ}}v^Q&=&\left(-K^I_{\phantom{I}P}K^J_{\phantom{J}Q}+K^J_{\phantom{J}P}K^I_{\phantom{I}Q}\right)v^Q\;.
\end{eqnarray}
By contracting equation \eqref{gauss5} with $v^P$, the result is equation \eqref{gauss4}. Thus, all information we need is actually contained in \eqref{gauss5}. Now, in order to suppress all free indices and solve for $v^0$, we contract \eqref{gauss5} with the two-dimensional curvature tensor itself. Hence, 
\begin{equation}\label{diffadm}
v^0=-\frac{\mathcal{R}_{IJ\phantom{P}0}^{\phantom{IJ}P}\left(2K^I_{\phantom{I}P}K^J_{\phantom{J}Q}+\mathcal{R}^{IJ}_{\phantom{IJ}{PQ}}\right)}{\mathcal{R}_{IJ\phantom{P}0}^{\phantom{IJ}P}\mathcal{R}^{IJ}_{\phantom{IJ}{P0}}}v^Q\;,
\end{equation}
characterizing the constrained diffeomorphism symmetry of the spacetime foliation. This ends the analysis of the consistency of the constraint \eqref{rel2} for vanishing torsion and general curvature within spacetime foliation $\mathbb{M}^3=\mathbb{M}^{2}\times\mathbb{R}_t$ for Riemannian geometry.

\begin{itemize}
 \item[(II)]{\textbf{On-shell curvature and general torsion}}
\end{itemize}

Still at the Poincaré symmetry, the second non-trivial situation we explore is to enforce the vanishing of the curvature while imposing the second condition of \eqref{rel2} for the torsion 2-form. Torsion now remains arbitrary. This non-standard condition ($R^{ij}=0$ and $i_vT^i=0$ for $T^i\ne0$) is a teleparallel gravity situation where gravity is described in a Weitzenböck manifold instead of a Riemannian one, see for instance \cite{MuellerHoissen:1983vc,DeAndrade:2000sf,Aldrovandi:2013wha}. In this case, the broken diffeomorphism group is now given by
\begin{equation}\label{sub2dif}
\mathrm{Diff}(2)_{\mathbb{M}^2}=\{v\in \mathrm{Diff}(3)\;\big|\;\mathbb{M}^3=\mathbb{M}^2\times\mathbb{R}_t\;;\;\;R^{ij}=0\;\;;\;\;i_vT^{i}=0\}\;.
\end{equation}

To find the condition over the components $v^\mu$ due to the constraints \eqref{rel2} we contract equation \eqref{gausstors} with $v^\mu$ to obtain
\begin{eqnarray}\label{gausstors1}
\mathcal{T}^{I}_{\phantom{I}\mu\nu}v^\nu&=&e^I_{\phantom{I}\alpha}T^{\alpha}_{\phantom{\alpha}\mu\nu}v^\nu+\Omega^I_{\phantom{I}\mu\nu}v^\nu\;.
\end{eqnarray}
The second equation of \eqref{rel2} implies that the first term of the \textit{rhs} in \eqref{gausstors1} vanishes, hence,
\begin{eqnarray}\label{gausstors2}
\mathcal{T}^{I}_{\phantom{I}\mu\nu}v^\nu&=&\Omega^I_{\phantom{I}\mu\nu}v^\nu\;.
\end{eqnarray}
The equation \eqref{gausstors2} states that $v^\mu$ twists when parallel transported along a spacelike closed curve and that the extrinsic torsion $\Omega$ enforces that $v^\mu$ will not end up at the same point. Now, splitting \eqref{gausstors2} in components, we find
\begin{eqnarray}\label{gausstors3}
\mathcal{T}^{I}_{\phantom{I}P0}v^P&=&0\;,
\end{eqnarray}
and
\begin{eqnarray}\label{gausstors4}
\mathcal{T}^{I}_{\phantom{I}P0}v^0&=&\left(\Omega^I_{\phantom{I}PQ}-\mathcal{T}^{I}_{\phantom{I}PQ}\right)v^Q\;.
\end{eqnarray}
Clearly, the contraction of the equation \eqref{gausstors4} with $v^P$ reproduces equation \eqref{gausstors3}. Thus, all relevant information contained in equation \eqref{gausstors2} lies at equation \eqref{gausstors4}. Then, taking the trace of equation \eqref{gausstors4} we finally obtain
\begin{equation}\label{diffadm1}
v^0=\frac{\left(\Omega-\mathcal{T}\right)^I_{\phantom{I}IQ}}{\mathcal{T}^{I}_{\phantom{I}I0}}v^Q\;,
\end{equation}
which is the constraint over the diffeomorphisms parameters we were seeking for. This ends the analysis of the consistency of the constraint \eqref{rel2} for vanishing curvature and general torsion within spacetime foliation $\mathbb{M}^3=\mathbb{M}^{2}\times\mathbb{R}_t$ for Weitzenböck geometry. 

\begin{itemize}
 \item[(III)]{\textbf{General torsion and general curvature}}
\end{itemize}

Now we consider both general constraints \eqref{rel2}. The broken Weitzenböck group is now given by definition \eqref{sub1}. In this case, all we have to do is to combine \eqref{diffadm} and \eqref{diffadm1} to obtain
\begin{eqnarray}\label{diffgen}
\left[\dfrac{\mathcal{R}_{IJ\phantom{P}0}^{\phantom{IJ}P}\left(2K^I_{\phantom{I}P}K^J_{\phantom{J}Q}+\mathcal{R}^{IJ}_{\phantom{IJ}{PQ}}\right)}{\mathcal{R}_{IJ\phantom{P}0}^{\phantom{IJ}P}\mathcal{R}^{IJ}_{\phantom{IJ}{P0}}}+\dfrac{\left(\Omega-\mathcal{T}\right)^I_{\phantom{I}IQ}}{\mathcal{T}^{I}_{\phantom{I}I0}}\right]v^Q=0,
\end{eqnarray}
which is the condition over $v^Q$ due to the chosen geometry. Clearly, in this case, the geometry is a Riemann-Cartan one since we are considering non-vanishing torsion and curvature. This ends the analysis of the consistency of the constraints \eqref{rel2} within spacetime foliation $\mathbb{M}^3=\mathbb{M}^{2}\times\mathbb{R}_t$ for Riemann-Cartan geometry.

\subsubsection{de Sitter symmetry}\label{3dconst1}

For de Sitter gauge symmetry, we have to study the constraints \eqref{rel4}. The procedure is pretty much the same as the one developed in Subsec.~\ref{3dconst}.

\begin{itemize}
 \item[(I)]{\textbf{On-shell torsion and general curvature}}
\end{itemize}
Again, on-shell torsion means that $T^i=0$ while general curvature assumes that $R^{ij}\ne \Lambda^ 2e^ie^j$ while $i_v(R^{ij}-\Lambda^ 2e^ie^j)=0$. Cleary, this situation is of a Riemannian geometry gravity. Therefore, the diffeomorphism group is broken to
\begin{equation}\label{sub1difa}
\mathrm{Diff}(2)_{\mathbb{M}^2}=\{v\in \mathrm{Diff}(3)\;\big|\;\mathbb{M}^3=\mathbb{M}^2\times\mathbb{R}_t\;;\;\;T^i=0\;\;;\;\;i_v\left(R^{ij}-\Lambda^2e^ie^j\right)=0\}\;.
\end{equation}

To find the meaning of the constraint \eqref{rel4} in terms of the diffeomorphism parameters $v^\mu$, we first project \eqref{rel4} in $\mathbb{M}^2$ to find
\begin{equation}\label{rel7}
\left[e^I_{\phantom{I}\alpha}e^J_{\phantom{J}\beta}R^{\alpha\beta}_{\phantom{\alpha\beta}{\mu\nu}}-\Lambda^2\left(e^{I}_{\phantom{I}\alpha}e^{\alpha}_{\phantom{\alpha}\mu}e^{J}_{\phantom{J}\beta}e^{\beta}_{\phantom{\beta}\nu}-e^{I}_{\phantom{I}\alpha}e^{\alpha}_{\phantom{\alpha}\nu}e^{J}_{\phantom{J}\beta}e^{\beta}_{\phantom{\beta}\mu}\right)\right]v^\nu=0\;.
\end{equation}
Hence, combining \eqref{rel7} with \eqref{gauss2}, we get
\begin{eqnarray}\label{gauss3}
\mathcal{R}^{IJ}_{\phantom{IJ}\mu\nu}v^\nu&=&\left[-K^I_{\phantom{I}\mu}K^{J}_{\phantom{J}\nu}+K^J_{\phantom{J}\mu}K^I_{\phantom{I}\nu}+\Lambda^2\left(\delta^{I}_{\phantom{I}\mu}\delta^{J}_{\phantom{J}\nu}-\delta^{I}_{\phantom{I}\nu}\delta^{J}_{\phantom{J}\mu}\right)\right]v^\nu\;.
\end{eqnarray}
Decomposing space and time we are lead to
\begin{eqnarray}\label{gauss4a}
\mathcal{R}^{IJ}_{\phantom{IJ}{P0}}v^P&=&0\;,
\end{eqnarray}
and
\begin{eqnarray}\label{gauss5a}
\mathcal{R}^{IJ}_{\phantom{IJ}{P0}}v^0+\mathcal{R}^{IJ}_{\phantom{IJ}{PQ}}v^Q&=&\left[-K^I_{\phantom{I}P}K^J_{\phantom{J}Q}+K^J_{\phantom{J}P}K^I_{\phantom{I}Q}+\Lambda^2\left(\delta^{I}_{\phantom{I}P}\delta^{J}_{\phantom{J}Q}-\delta^{I}_{\phantom{I}Q}\delta^{J}_{\phantom{J}P}\right)\right]v^Q\;.
\end{eqnarray}
Expression \eqref{gauss4a} is exactly the same as \eqref{gauss4}. The novelty is equation \eqref{gauss5a} where the purely spatial vector field $v^Q$ is affected by the extrinsic curvature and by the cosmological constant term when parallel transported along an infinitesimal spatial closed path. The explicit constraint over the diffeomorphism parameters is easily found: First, one saturates the free indices in \eqref{gauss5a} by contracting it with the leaf curvature. Then, one solves the equation for $v^0$, namely, 
\begin{equation}\label{diffadm2}
v^0=\dfrac{\mathcal{R}_{IJ\phantom{P}0}^{\phantom{IJ}P}\left[-2K^I_{\phantom{I}P}K^J_{\phantom{J}Q}-\mathcal{R}^{IJ}_{\phantom{IJ}{PQ}}+2\Lambda^2\delta^{I}_{\phantom{I}P}\delta^{J}_{\phantom{J}Q}\right]v^Q}{\mathcal{R}_{IJ\phantom{P}0}^{\phantom{IJ}P}\mathcal{R}^{IJ}_{\phantom{IJ}{P0}}}\;,
\end{equation}
where the influence of the cosmological constant is evident. This ends the analysis of the consistency of the constraint \eqref{rel4} for vanishing torsion and general curvature within spacetime foliation $\mathbb{M}^3=\mathbb{M}^{2}\times\mathbb{R}_t$ for Riemannian geometry.

\begin{itemize}
 \item[(II)]{\textbf{On-shell de Sitter curvature and general torsion}}
\end{itemize}
This case is similar to the case of on-shell de Sitter curvature and general torsion for Poincaré symmetry because we have $R^{ij}=\Lambda^2e^ie^j$ fixed instead of $R=0$ (See Subsec.~\ref{3dconst} item (II)). Therefore the constraint is the exactly same that \eqref{diffadm1}. Nevertheless, while in the Poincaré symmetry case we were in a Weitzenböck scenario, now we are at a Riemann-Cartan scenario since the vacuum curvature does not vanish anymore. The broken diffeomorphism group is also different from \eqref{sub2dif},
\begin{equation}\label{sub1.1aa}
\mathrm{Diff}(2)_{\mathrm{fol}}=\{v\in \mathrm{Diff}(3)\;\big|\;R^{ij}-\Lambda^2e^ie^j=0\;\;;\;\;i_vT^i=0\}\;.
\end{equation}

This ends the analysis of the consistency of the constraint \eqref{rel4} for general torsion and de Sitter vacuum curvature within spacetime foliation $\mathbb{M}^3=\mathbb{M}^{2}\times\mathbb{R}_t$ for Riemann-Cartan geometry.

\begin{itemize}
 \item[(III)]{\textbf{General torsion and general curvature}}
\end{itemize}

To find the general constraint over $v^\mu$ when considering general curvature and torsion in the constraints \eqref{rel4} is a simple task. The result follows from the direct combination of the the previous results \eqref{diffadm2} and \eqref{diffadm1},
\begin{eqnarray}\label{diffgen1}
\left[\dfrac{\mathcal{R}_{IJ\phantom{P}0}^{\phantom{IJ}P}\left(-2K^I_{\phantom{I}P}K^J_{\phantom{J}Q}-\mathcal{R}^{IJ}_{\phantom{IJ}{PQ}}+2\Lambda^2\delta^{I}_{\phantom{I}P}\delta^{J}_{\phantom{J}Q}\right)}{\mathcal{R}_{IJ\phantom{P}0}^{\phantom{IJ}P}\mathcal{R}^{IJ}_{\phantom{IJ}{P0}}}-\dfrac{\left(\Omega-\mathcal{T}\right)^I_{\phantom{I}IQ}}{\mathcal{T}^{I}_{\phantom{I}I0}}\right]v^Q=0\;.
\end{eqnarray}
The broken diffeomorphic group is, obviously, the one defined in \eqref{sub1a}. Also, from the general nature of torsion and curvature, we are again in a Riemann-Cartan gravity scenario. This ends the analysis of the consistency of the constraint \eqref{rel4} for general curvature and torsion within spacetime foliation $\mathbb{M}^3=\mathbb{M}^{2}\times\mathbb{R}_t$ for a Riemann-Cartan geometry.

\subsection{Four dimensions}

It is not difficult to see that the constraints \eqref{rel9} are essentially the same as \eqref{rel4}, except for the cosmological constant factor and the spacetime dimension. Thus, we follow exactly the same line of reasoning exposed in Sec.~\ref{ads3d}, but for the foliated spacetime $\mathbb{M}^4=\mathbb{M}^3\times\mathbb{R}_t$. Moreover, we avoid the use of the word "on-shell" in this part because we will not deal with any on-shell case. Other differences will also appear.

\begin{itemize}
 \item[(I)]{\textbf{Vanishing torsion and general curvature}}
\end{itemize}

By assuming vanishing torsion and a generic curvature, the broken diffeomorphism group now reads 
\begin{equation}\label{sub5}
\mathrm{Diff}(3)_{\mathrm{fol}}=\left\{v\in \mathrm{Diff}(4)\;\big|\;T^a=0\;\;;\;\;i_v\left(R^{ab}-\dfrac{\Lambda^2}{3}e^ae^b\right)=0\right\}\;.
\end{equation}
The scenario is a Riemannian gravity. The consequence is that the constraints \eqref{rel9} are reduced to
\begin{eqnarray}\label{gauss6}
\mathcal{R}^{AB}_{\phantom{AB}\mu\nu}v^\nu&=&\left[-K^A_{\phantom{A}\mu}K^{B}_{\phantom{B}\nu}+K^B_{\phantom{B}\mu}K^A_{\phantom{A}\nu}+\dfrac{\Lambda^2}{6}\left(\delta^{A}_{\phantom{A}\mu}\delta^{B}_{\phantom{B}\nu}-\delta^{A}_{\phantom{A}\nu}\delta^{B}_{\phantom{B}\mu}\right)\right]v^\nu\;,
\end{eqnarray}
where $\{A,B\ldots=1,2,3\}$ are spatial indices of the manifold $\mathbb{M}^3$ and where Greek indices vary now as $\{0,1,2,3\}$ characterizing the world indices of the manifold $\mathbb{M}^4$. The constraint over the diffeomorphism parameter could be found according to the previous case (See Subsec.~\eqref{3dconst1} item (I)): First, we split the indices in \eqref{gauss6} into space and time. To saturate the remaining three free indices we now have the three-dimensional Levi-Civita simbol at our disposal\footnote{In the three-dimensional case we had also three free indices but no three-index Levi-Civita symbol.}, resulting in 
\begin{equation}\label{diffadm2a}
v^0=-\dfrac{\epsilon_{AB}^{\phantom{AB}P}\mathcal{R}^{AB}_{\phantom{AB}{PQ}}}{\epsilon_{AB}^{\phantom{AB}P}\mathcal{R}^{AB}_{\phantom{AB}{P0}}}v^Q\;,
\end{equation}
which is much more simple than the result found in \eqref{diffadm2} and, remarkably, does not depend explicitly on the cosmological constant. This ends the analysis of the consistency of the constraint \eqref{rel9} for vanishing torsion and general curvature within spacetime foliation $\mathbb{M}^4=\mathbb{M}^{3}\times\mathbb{R}_t$ for a Riemannian geometry.

\begin{itemize}
 \item[(II)]{\textbf{General torsion and de Sitter curvature}}
\end{itemize}

Now we set the de Sitter curvature $R^{ab}=\dfrac{\Lambda^2}{3}e^ae^b$ and general torsion. The broken diffeomorphic group is now given by
\begin{equation}\label{sub2adif}
\mathrm{Diff}(3)_{\mathbb{M}^3}=\left\{v\in \mathrm{Diff}(4)\;\big|\;\mathbb{M}^3=\mathbb{M}^2\times\mathbb{R}_t\;;\;\;R^{ab}-\dfrac{\Lambda^2}{3}e^ae^b=0\;\;;\;\;i_vT^{a}=0\right\}\;.
\end{equation}
And, for the diffeomorphisms parameters, we get 
\begin{equation}\label{diffadm1c}
v^0=\frac{\left(\Omega-\mathcal{T}\right)^A_{\phantom{A}AQ}}{\mathcal{T}^{A}_{\phantom{A}A0}}v^Q\;,
\end{equation}
such as encountered in \eqref{diffadm1}. 

\begin{itemize}
 \item[(III)]{\textbf{General torsion and general curvature}}
\end{itemize}

Finally, for the general torsion and curvature, we combine \eqref{diffadm2a} and \eqref{diffadm1c} to find
\begin{equation}\label{diffgen2}
\left[\dfrac{\epsilon_{AB}^{\phantom{AB}P}\mathcal{R}^{AB}_{\phantom{IJ}{PQ}}}{\epsilon_{AB}^{\phantom{AB}P}\mathcal{R}^{AB}_{\phantom{AB}{P0}}}+\dfrac{\left(\Omega-\mathcal{T}\right)^A_{\phantom{A}AQ}}{\mathcal{T}^{A}_{\phantom{A}A0}}\right]v^Q=0\;.
\end{equation}
Remarkably, the cosmological constant has no influence on this constraint. The reason is that we have used the Levi-Civita symbol to saturate the indices instead of the leaf curvature. Finally, the broken diffeomorphism group is the generic one given by expression \eqref{sub1b}.

\section{Principal bundles, gauge theories and gravity}\label{fb}

In this section we provide a quick mathematical analysis of our results. Specifically, we discuss the mapping of a principal bundle (gauge theory) into a coframe bundle (gravity as geometrodynamics) from the point of view of fiber bundle theory \cite{Kobayashi:1963fg,Nakahara:2003nw,Bertlmann:1996xk,Singer:1978dk,Daniel:1979ez,Bleecker:1981me}. This analysis serve as a mathematical support of the previous sections.

The principal bundle for which a gauge connection is generated \cite{Daniel:1979ez} is denoted by $\{P,\mathbb{G},\mathbb{M}^n,\pi\}$ where $\pi:{P}\longmapsto\mathbb{M}^n$. In short notation, we write simply ${P}(\mathbb{M}^n,{\mathbb{G}})$. The fiber and the structure group are both a Lie group $\mathbb{G}$. The base space is a $n$-dimensional manifold $\mathbb{M}^n$, often recognized as the spacetime. The total space is the non-trivial product $P=\mathbb{M}^n\times\mathbb{G}$. The projection map $\pi$ is a continuous surjective map and the local triviality condition of $P $ is guaranteed by the homeomorphism $\phi_i:{\pi}^{-1}(\{\mathbb{M}^n\}_i)\longmapsto\{\mathbb{M}^n\}_i\times\mathbb{G}$, where $\{\mathbb{M}^n\}_i$ are open sets covering $\mathbb{M}^n$. The definition of $P$ can be regarded as a formal way to describe the localization of the Lie group $\mathbb{G}$ over the $n$-dimensional spacetime, associating to each point  $x\in\mathbb{M}^n$ different values for the elements $u(x)\in\mathbb{G}$. The gauge connection appears in the definition of parallel transport in $P$. Gauge transformations are associated with coordinates changes in $P$ by keeping the coordinates of the base space fixed, i.e., a translation along the fiber ${\pi}^{-1}$. Furthermore, for any connection 1-form $A$ there is a curvature 2-form defined over $P$ given by $F=dA+AA$ which is recognized as the field strength in gauge theories.

Gravity theories can also be described in terms of principal bundles and, consequently, as gauge a theory. Initially, we will take a manifold\footnote{We included a "tilde" in the manifold of gravity to emphasize that this manifold is not the manifold without the "tilde" of the gauge theory. This distinction was not made before because there was no need until now.} $\widetilde{\mathbb{M}}^n$ and the respective collection of tangent spaces $T_X(\widetilde{\mathbb{M}}^n)$ at all points $X\in\widetilde{\mathbb{M}}^n$. The collection of all tangent spaces defines the tangent bundle $\mathbb{T}$. In this bundle, $\widetilde{\mathbb{M}}^n$ is the base space, the fiber is denoted by $T_X(\widetilde{\mathbb{M}}^n)$ and the structure group is the $GL(n,\mathbb{R}^n)$. Similarly the cotangent bundle $\mathbb{T}^{\ast}$ is defined as union of all cotangent spaces in $\widetilde{\mathbb{M}}^n$. The fundamental structure for gravity theories is, for instance, the coframe bundle. The coframe bundle is associated with the cotangent bundle. A typical fiber at $X\in\widetilde{\mathbb{M}}^n$ is the set of all local coframes, which are represented by vielbeins $e$ defined in $T^\ast_X (\mathbb{M}^n)$. Consequently, given a fixed coframe at $T^\ast_X (\widetilde{\mathbb{M}}^n)$, there will be an infinite number of equivalent coframes related to the former by transformations of $GL(n,\mathbb {R}^n)$. Obviously, the fibers coincide with the group. We can define the coframe bundle as $P^\ast=(GL(n,\mathbb{R}^n),\widetilde{\mathbb{M}}^n)$. Note that $GL(n,\mathbb{R}^n)$ can be trivially reduced to orthogonal group $SO(n)$ which leads to bundle $P'^\ast (SO(n),\widetilde{\mathbb{M}}^n)$ that coincides with the mathematical structure of Einstein-Cartan gravity. See \cite{Nash:1983cq,Sobreiro:2010ji}. In $P'^\ast$ the $SO(n)$ algebra-valued connection is the spin-connection 1-form $\omega$. The curvature 2-form in $P'^\ast$ is defined as $R=d\omega+\omega\omega$ while the torsion 2-form is $T=de+\omega e$. It is useful to understand that the only difference between the coframe bundle $P'^\ast$ and the gauge principal bundle $P$ is that the structure group (and fiber) $SO(n)$ is also the group of local isometries. Hence, gravity can be visualized as a gauge theory where the gauge group is identified with the spacetime local isometries.

In order to connect a gauge theory with a gravity theory in the light of the principal bundles above described, let us start with the Poincaré three-dimensional case. First, we assume that the original Chern-Simons action \eqref{cs1} is constructed over a flat manifold, say, the three dimensional Euclidean spacetime. It is possible then to reinterpret the transition from action \eqref{cs1} to \eqref{eh1} as a kind of duality between a Chern-Simons action in Euclidean spacetime and a gravity action over a general manifold. The action \eqref{cs1} is a standard Chern-Simons gauge theory and, in principle, has nothing to do with gravity. Essentially, the field $A$ is a connection over the principal bundle $(SU(2)\times\mathbb{R}^3,\mathbb{R}^3)$ where the base space is the three dimensional Euclidean space $\mathbb{R}^3$ and the structure group (and fiber) is the gauge group. On the other hand, action \eqref{eh1} is a gravity action describing the geometrodynamics of a three-dimensional manifold $\widetilde{\mathbb{M}}^3$. Thus, the spin-connection $\omega$ is a connection over a coframe bundle $(SO(3),\widetilde{\mathbb{M}}^3)$ where the base space is $\widetilde{\mathbb{M}}^3$, the structure group is $SO(3)$ and a typical fiber at a point $X\in\widetilde{\mathbb{M}}^3$ is the collection of all coframes that can be defined at the tangent space $T_X(\widetilde{\mathbb{M}}^3)$. Obviously the coframes are identified with the dreibein $e$.

In \cite{Sobreiro:2011hb}, a map between a gauge theory over a Euclidean spacetime and a geometrodynamics gravity theory was proposed in four dimensions. However, an Inönü-Wigner contraction and symmetry breaking were required. In the present case, neither of these effects is needed. In fact, the formal map $f:(SU(2)\times\mathbb{R}^3,\mathbb{R}^3)\longmapsto(SO(3),\widetilde{\mathbb{M}}^3)$ can be understood as a set of maps $f_i$ as follows: An isomorphism between base spaces $f_0:\mathbb{R}^3\longmapsto\widetilde{\mathbb{M}}^3$. The $SU(2)$ sector of the gauge group is mapped into the local isometry group $SO(3)$ of $\widetilde{\mathbb{M}}^3$, while the gauge sector $\mathbb{R}^3$ is mapped into the diffeomorphisms group, namely, $f_1: SU(2)\longmapsto SO(3)$ and $f_2: \mathbb{R}^3\longmapsto \mathrm{Diff}(3)$, respectively. To $f_2$ be properly defined, one has to impose the relations \eqref{rel1}. We also demand that the space of \emph{p}-forms in $\mathbb{R}^3$ is identified with the space of \emph{p}-forms in $\widetilde{\mathbb{M}}^3$, namely $\Omega^p$ and $\widetilde{\Omega}^p$, by means of $f_3:\Omega^p\longmapsto\widetilde{\Omega}^p$. Finally, by imposing the relation \eqref{id1} as an extra condition to $f_3$, the coframe bundle $(SO(3),\widetilde{\mathbb{M}}^3)$ can be constructed from $(SU(2)\times\mathbb{R}^3,\mathbb{R}^3)$.

It is important to stress that $f_0$ must be an isomorphism, otherwise ambiguities would appear.  To see this, let us suppose that $f_0$ has not a one-to-one correspondence. Then, two fibers at point $x_1\in\mathbb{R}^3$ and $x_2\in\mathbb{R}^3$ can be identified with one fiber at $X\in\widetilde{\mathbb{M}}^3$. Clearly, this is not possible because the two original fibers are supposed to be independent from each other. Moreover, the explicit form of $f_0$ can be straightforwardly derived for any spacetime dimension, see\cite{Sobreiro:2011hb},
\begin{equation}
L^{\nu}_{\phantom{\nu}\mu}=\left(\dfrac{\tilde{g}}{g}\right)^{\sfrac{1}{2n}}g^{\nu\alpha}\tilde{g}_{\alpha\mu},\label{matriz}
\end{equation}
where $L$ is the matrix that transforms the components of forms $\Omega^p$ into forms $\widetilde{\Omega}^p$; $g=|\det g_{\mu\nu}|$ and $\tilde{g}=|\det\tilde{g}_{\mu\nu}|$; $g_{\mu\nu}$ e $\tilde{g}_{\mu\nu}$ are respectively original and target metrics. Obviously, for the present case we have $g_{\mu\nu}=\delta_{\mu\nu}$ and $n=3$. Thus,
\begin{equation}
L^{\nu}_{\phantom{\nu}\mu}=\left(\dfrac{1}{g}\right)^{\sfrac{1}{6}}g^{\nu\alpha}\tilde{g}_{\alpha\mu}.\label{matriz1}
\end{equation}

Another important remark is that the map $f_2$ is obtained from specific values of the parameters which are functions of $e$ and $\omega$. This means that $f_2$ is not a direct equivalence between $\mathbb{R}^3$ and $\mathrm{Diff}(3)$, but a special case where \eqref{rel1} is imposed. Remarkably, all geometrical properties of the principal bundle $(SU(2)\times\mathbb{R}^3,\mathbb{R}^3)$ can be used to construct the coframe bundle $(SO(3),\widetilde{\mathbb{M}}^3)$, including the diffeomorphic transition functions.

In the case where the cosmological term is included, a similar map can be defined. The difference with respect to the previous case is that $f_2:S(3)\longmapsto \mathrm{Diff}(3)$. In this case, \eqref{rel1} has a stronger appeal because $f_2$ generate diffeomorphisms from a non-Abelian sector of the gauge group which is not even a subgroup, but a symmetric coset space. 

Since the starting manifold is an Euclidean space, the target manifold is a deformed space whose geometric properties are defined through
\begin{eqnarray}\label{geom0}
g_{\mu\nu}&=&\delta_{ij}\theta^i_\mu \theta^j_\nu\;,\nonumber\\
\Gamma^\alpha_{\mu\nu}&=&\delta_{ij}g^{\alpha\beta}\theta^j_\beta\left(\partial_\mu \theta_\nu^i+A^i_{\mu k}\theta^k_\nu\right)\;.
\end{eqnarray}
In that sense, the gauge field is visualized as "absorbed" by the spacetime itself. This "absorption" deforms the base space and generate the effective geometry.

Once again, we remark that the identification \eqref{id1} is only possible if a mass parameter is at our disposal. In three dimensions, we have used the coupling constant $\kappa$. Since this parameter is inherent to the theory, this Chern-Simons-gravity duality can be performed at any scale of the theory, a feature that is not possible in four dimensions. Moreover, it is important to be understood that the action \eqref{cs1} is a theory where the gauge symmetry has nothing to do with the geometrical properties of the spacetime. Only after the mapping, the gauge symmetry is identified with the geometrical properties of the base space and the fields with the geometrical objects that describe the dynamics of this geometry. 

The previous discussion can easily be extended to the four-dimensional case. We assume that the Yang-Mills action \eqref{ym1} is constructed over a flat four-dimensional Euclidean manifold. The transition from action \eqref{ym1} to \eqref{ym2} can be taken as a duality between a Yang-Mills action in Euclidean spacetime and a gravity action over a general manifold. The field $Y$ is a connection over the principal bundle $(SO(5),\mathbb{R}^4)$. The gravity action \eqref{ym2} describes the geometrodynamics of a four-dimensional manifold $\widetilde{\mathbb{M}}^4$. Thus, the spin-connection $\omega$ is a connection over a coframe bundle $(SO(4),\widetilde{\mathbb{M}}^4)$. Again, the concepts of Inönü-Wigner contraction and symmetry breaking are not required. The formal map $f:(SO(5),\mathbb{R}^4)\longmapsto(SO(4),\widetilde{\mathbb{M}}^4)$ is a set of maps $f_i$ as follows: An isomorphism between base spaces $f_0:\mathbb{R}^4\longmapsto\widetilde{\mathbb{M}}^4$. The $SO(4)$ sector of the gauge group is mapped into the local isometry group $SO(4)$ of $\widetilde{\mathbb{M}}^4$, while the gauge sector $S(4)=SO(5)/SO(4)$ is mapped into the diffeomorphisms group, namely, $f_1: SO(4)\longmapsto SO(4)$ and $f_2: S(4)\longmapsto \mathrm{Diff}(4)$, respectively. Moreover, the space of \emph{p}-forms in $\mathbb{R}^4$ is identified with the space of \emph{p}-forms in $\widetilde{\mathbb{M}}^4$ by means of $f_3:\Omega^p\longmapsto\widetilde{\Omega}^p$. Finally, by imposing the relation \eqref{id3} to $f_3$, the coframe bundle $(SO(4),\widetilde{\mathbb{M}}^4)$ is constructed from the principal bundle $(SO(5),\mathbb{R}^4)$. We stress out that $f_0$ must be an isomorphism due to the same reasons of the three-dimensional case. Also, $f_2$ is not a direct equivalence between $S(4)$ and $\mathrm{Diff}(4)$, but a special case where \eqref{rel8} is imposed. Once again, we remark that the identification \eqref{id3} is only possible if a mass parameter is at our disposal. In four dimensions (in contrast to the three-dimensional case), the action \eqref{ym1} is massless. Hence, the mapping depends whenever the theory develops mass parameters. Fortunately, Yang-Mills theories develop a few non-perturbative effects that lead to the appearance of mass scales \cite{Dudal:2005na,Dudal:2011gd,Capri:2015ixa,Capri:2016aqq,Capri:2016aif}. Thus, the mapping would only hold for low a low energy regime.

To end this section, we wish to recall that alternative mappings that could be applied in the same context can be found in\cite{Obukhov:1998gx} and references therein.

\section{Conclusions}\label{concl}

In this work, we were able to provide a reinterpretation of Witten's three-dimensional equivalence result \cite{Witten:1988hc} by showing how one can construct an equivalent gravity theory originated from a Chern-Simons theory in Euclidean spacetime. Such equivalence required a mass parameter (coupling parameter) and a set of geometric constraints (\eqref{rel2} for Poincaré gauge theory and; \eqref{rel4} for de Sitter gauge theory). The method is valid at any scale since the coupling parameter already carries mass dimension. The method was also applied to the four dimensional case for the Yang-Mills action with $SO(5)\equiv SO(4)\times S(4)$ gauge symmetry and considering flat spacetime. In this case, the coupling parameter is dimensionless and another parameter is required. Fortunately, there are a few mass parameters at our disposal in four-dimensional Yang-Mills theories at lower energy regime. For instance, the Gribov parameter is a nice candidate due to the recent discovery that it is a gauge invariant quantity \cite{Capri:2015ixa,Capri:2016aqq,Capri:2016aif}. A set of geometric constraints were also required, namely \eqref{rel9}. In both cases (Three and four dimensions), the gauge field can be associated with geometrical objects, namely, the vielbein and spin-connection, while the gauge group can be identified with local isometries and diffeomorphisms. Hence, the gauge field degrees of freedom can be somehow seen as being absorbed by spacetime. It is worth mentioning that the present method is comparable with the mechanism described in\cite{Sobreiro:2011hb,Assimos:2013eua,Sobreiro:2016fks} for Yang-Mills theory. The advantage here is the fact that we do not depend on the running of the mass parameters to perform the mapping nor on an Inönü-Wigner contraction mechanism. The present mechanism merely depends on the existence of a mass parameter and not on its behavior with respect to the energy scale.

We have also argued that the constraints \eqref{rel2}, \eqref{rel4} and \eqref{rel9} defines foliations \cite{Dufour:2005th,Lavau:2018th} with a restricted class of diffeomorphisms in the foliated manifold. For physical reasons, we have performed the study of such foliations in a scenario where the leafs are spacelike and evolve in time. Nevertheless, the method could be generalized to any kind of foliation. There are a few ways to fulfill the referred constraints. In three dimensions we have four main cases: 
\begin{itemize}
\item The usual Witten on-shell conditions which lead to gravity theories displaying full diffeomorphism symmetry. However, vacuum geometry must be imposed.
\item On-shell torsion and general curvature: The result is a Riemannian Einstein-Hilbert gravity (\eqref{eh1} for Poincaré symmetry and \eqref{eh2} for de Sitter symmetry) with restricted diffeomorphism symmetry, see \eqref{diffadm} for the Poincaré symmetry and \eqref{diffadm2} for the de Sitter symmetry.
\item On-shell curvature and general torsion: The result is a Weitzenböck Einstein-Hilbert gravity (\eqref{eh1} for Poincaré symmetry and \eqref{eh2} for de Sitter symmetry) with restricted diffeomorphism symmetry, see \eqref{diffadm1} for both, Poincaré and de Sitter symmetries.
\item General torsion and curvature: The result is a Riemann-Cartan Einstein-Hilbert gravity (\eqref{eh1} for Poincaré symmetry and \eqref{eh2} for de Sitter symmetry) with restricted diffeomorphism symmetry, see \eqref{diffgen} for the Poincaré symmetry and \eqref{diffgen1} for the de Sitter symmetry.
\end{itemize} 
In four dimensions there is no room for the on-shell case and we are lead to three main cases: 
\begin{itemize}
\item Vanishing torsion and general curvature: The result is a Riemannian gravity (action \eqref{ym2} with $T=0$) with restricted diffeomorphism symmetry, see \eqref{diffadm2a}.
\item de Sitter curvature and general torsion: The result is a Riemann-Cartan gravity (action \eqref{ym2} with curvature given by $R^{ab}=\dfrac{\Lambda^2}{3}e^ae^b$) with restricted diffeomorphism symmetry, see \eqref{diffadm1c}.
\item General torsion and curvature: The result is a Riemann-Cartan Einstein-Hilbert gravity (action \eqref{eh1}) with restricted diffeomorphism symmetry, see \eqref{diffgen2}.
\end{itemize} 

A mathematical analysis in terms of fiber bundle theory were also performed. This study allows the comprehension of how a principal bundle describing a classical gauge theory can be mapped into a coframe bundle describing a classical gravity theory.

Finally, we wish to point out two possibly interesting analysis that are left for future investigation: First, the quantum consistency of the mechanism. In three dimensions, this may be helped by the fact that Chern-Simons theories are finite\cite{Delduc:1990je,Piguet:1995er}. It is appropriate to dedicate a few words about the 3D quantum gravity scenario explored in\cite{Carlip:1998uc} where the author scrutinize both, the canonical and functional quantization methods. The ADM decomposition, geometric structures, covariant phase space are studied in second and first order formalisms. See also\cite{Cotler:2018zff,Saghir:2017fol} for more recent developments. In four dimensions, $SO(5)$ Yang-Mills theories are known to be renormalizable and unitary\cite{Itzykson:1980rh}. Remembering that in both cases the constraints must be taken into account. Second, the inclusion of matter fields (bosonic and fermionic) may produce interesting effects to be studied. Possibly extra constraints may be required. It is clear then that the results of the present work generalizes the classical results of\cite{Carlip:1998uc} and opens the possibility to account for other spacetime foliations than the ADM decomposition. Moreover, as mentioned before, the whole setup is also generalized to four dimensions for the Yang-Mills theory.

\section*{Acknowledgements}

The authors are grateful to Conselho Nacional de Desenvolvimento Científico e Tecnológico, The Coordenação de Aperfeiçoamento de Pessoal de Nível Superior (CAPES) and the Pró-Reitoria de Pesquisa, Pós-Graduação e Inovação (PROPPI-UFF) are acknowledge for financial support.

\bibliography{BIB}
\bibliographystyle{unsrt}

\end{document}